\documentclass[graybox]{svmult}

\usepackage{type1cm}        
\usepackage{makeidx}         
\usepackage{graphicx}        
\usepackage{multicol}        
\usepackage[bottom]{footmisc}
\usepackage[inline]{enumitem}

\usepackage{newtxtext}       %
\usepackage[varvw]{newtxmath}       

\usepackage{algpseudocode}
\usepackage{algorithm}

\usepackage{xcolor}
\usepackage{listings}
\usepackage{fancyvrb}
\definecolor{dkgreen}{rgb}{0,0.6,0}
\definecolor{gray}{rgb}{0.5,0.5,0.5}
\definecolor{mauve}{rgb}{0.58,0,0.82}

\usepackage[binary-units]{siunitx}
\makeindex             

\begin{document}

\title*{A new framework for calibrating COVID-19 SEIR models with spatial-/time-varying coefficients
using genetic and sliding window algorithms }
\titlerunning{A new framework for calibrating COVID-19 SEIR model}
\author{Huan Zhou, Ralf Schneider}
\institute{Huan Zhou \at High Performance Computing Center Stuttgart (HLRS), \email{huan.zhou@hlrs.de}
\and Ralf Schneider \at High Performance Computing Center Stuttgart (HLRS), \email{ralf.schneider@hlrs.de}}
%
%
\maketitle

\abstract*{A susceptible-exposed-infected-removed (SEIR) model assumes spatial-/time-varying coefficients to model the effect of non-pharmaceutical interventions (NPIs) on the regional and temporal distribution of COVID-19 disease epidemics. A significant challenge in using such model is their fast and accurate calibration to observed data from geo-referenced hospitalized data, i.e., efficient estimation of the spatial-/time-varying parameters. In this work, a new calibration framework is proposed towards optimizing the spatial-/time-varying parameters of the SEIR model. We also devise a method for combing the overlapping sliding window technique (OSW) with a genetic algorithm (GA) calibration routine to automatically search the segmented parameter space. Parallelized GA is used to reduce the computational burden. Our framework abstracts the implementation complexity of the method away from the user. It provides high-level APIs for setting up a customized calibration system and consuming the optimized values of parameters. We evaluated the application of our method on the calibration of a spatial age-structured microsimulation model using a single objective function that comprises observed COVID-19-related ICU demand. The results reflect the effectiveness of the proposed method towards estimating the parameters in a changing environment.}

\abstract{
A susceptible-exposed-infected-removed (SEIR) model assumes spatial-/time-varying coefficients to model the effect of non-pharmaceutical interventions (NPIs) on the regional and temporal distribution of COVID-19 disease epidemics. A significant challenge in using such model is their fast and accurate calibration to observed data from geo-referenced hospitalized data, i.e., efficient estimation of the spatial-/time-varying parameters. In this work, a new calibration framework is proposed towards optimizing the spatial-/time-varying parameters of the SEIR model. We also devise a method for combing the overlapping sliding window technique (OSW) with a genetic algorithm (GA) calibration routine to automatically search the segmented parameter space. Parallelized GA is used to reduce the computational burden. Our framework abstracts the implementation complexity of the method away from the user. It provides high-level APIs for setting up a customized calibration system and consuming the optimized values of parameters. We evaluated the application of our method on the calibration of a spatial age-structured microsimulation model using a single objective function that comprises observed COVID-19-related ICU demand. The results reflect the effectiveness of the proposed method towards estimating the parameters in a changing environment.
}

\section{Introduction}
\label{hz:intro}
The susceptible-exposed-infected-removed (SEIR)~\cite{li1995global} model has found wide application in the realm of epidemiology, mostly for quantifying the transmission dynamics of infectious diseases with incubation. The standard SEIR model assumes that the transmissibility-related parameter (denoted as $\mu$) is time/space-invariant. However, the characteristics of an epidemic (especially the recent COVID-19 pandemic) suggest that this parameter impacts the basic reproduction number $R0$ and can vary, due to the season-forcing, the presence of a vaccine, the communicative connection between different regions out of the demands for the daily commute or visits and non-pharmaceutical interventions (NPIs), e.g., political measures. To better evaluate the trend of the COVID-19 pandemic, the SEIR model has consistently been extended by incorporating the time/region-varying characteristic into the parameter 
$\mu$~\cite{klusener2020forecasting,gleeson2022,girardi2023seir}.
Therefore, estimating the values of parameter $\mu$, i.e., calibration of these extended SEIR models to the observed data -- e.g., cumulative ICU cases of COVID-19, is a crucial technical challenge that is enhanced by the time/region-varying nature of the epidemic.

The parameter $\mu$ is often deemed to be a real-valued continuous variable. This, along with its characteristic of spatial variation, easily leads to high-dimensional and broad search space. Thus, evaluating the parameter $\mu$ 
at a certain point in time can even be seen as a large-scale optimization problem. Naturally, expensive computation
is inevitable for the practitioners to solve it.
Besides, there are always communicative connections between different locations out of the demands
for the daily commute or visits, i.e., parameters for different regions at certain time in point are not completely non-separable, but correlated to some extent because of the interactions between them. To this end, the values of the parameter $\mu$ for all locations (multi-dimensional parameter) should be estimated as a whole in the course of the calibration.
These points prompt an important question: which optimization method we should choose to efficaciously evaluate an ensemble of values
of $\mu$ in such a large-scale calibration problem? In our study, we employ Evolutionary Algorithm (EA), that can easily and efficiently be parallelized, due to the intuitive independence between individuals. 
In this regard, the calibration procedure has a chance to get greatly accelerated as long as the computational capacity is sufficient.
Moreover, a combination of all optimized values of parameter $\mu$ during a single EA run can be obtained, since it can manage problems with non-separable interactions, which aligns with the principle
of parameter interactions in the problem of our concern as well as many other real-world problems~\cite{yu2010introduction,iorio2006incorporating}.
There are two important EA method variants, Genetic Algorithm (GA) and Differential Evolution (DE).
DE involves a direct-based mutation operator, which can better refine the solution than GA from generation to generation. 
Traditionally, both of them necessitate fitness evaluation of individuals after a new generation of candidates is formed.
In many real-world optimization problems (including ours), the evaluation needs considerable computations;
thus we attempt to constraint the number of generations (i.e. reduce the number of fitness function evaluations)
for avoiding an extremely slow search process.
An interesting study~\cite{santos2012comparison} observed that 
the fitness value of the best individuals drops much 
faster in the initial generations of GA than in DE. This observation drives us to reveal a preference for GA over DE since
GA allows a good convergence within an affordable number of generations. The limited number 
of generations, however, speeds up the search process at the expense of degraded solution accuracy.
Certainly, further measures should be taken to rule out a heavy compromise on solution accuracy.
However, Parallel GAs (PGAs) are 
actually not frequently implemented according to the existing studies on the application of GAs to model calibration~\cite{leardi2003nature,leardi2001genetic,leardi2007genetic,dolan2022model,monteiro2020influence,guo2021calibrating,barnhart2017moesha}, 
two of which~\cite{guo2021calibrating,barnhart2017moesha} used dedicated distributed GAs written in Matlab and Python.
The merit of "ease-of-parallelization" fails to be fully revealed on high-performance computing clusters since
Matlab and Python are not friendly to computation with high requirements for time performance.
Moreover, variants of GA emerge with reinforcements with its operators, e.g. Elite GA (EGA,~\cite{guo2021calibrating}) enhances the selection operation with elite individuals.

Not surprisingly, continuously 
evaluating the values of parameter $\mu$ over time is even more complex and computationally expensive. 
As we mentioned before, the parameter $\mu$ is time-dependent, i.e., changes regularly.
If we engage in evaluating the sequence of values at one calibration execution, the number of optimized
parameters a GA works with is overwhelming and induces an extremely broad search space.
To enable a high probability of finding a near-optimal solution in a broad search space, the GA method demands a very high population.
In this regard, more computation capacity is needed to parallelize the fitness evaluation of individuals for 
keeping a fast search process.
However, the computation capacity is finite 
and not that easy to be fully obtained and reserved for a considerable time.
Therefore, evaluating the sequence of values at a one-off calibration execution is not realistic. Here another question is raised:
how can the problem of evaluating the values of such a spatial-/time-varying parameter be simplified? 
Actually, the ordered sequence of optimized values can be seen
as a stream of time series data~\cite{esling2012time}; thus problem partitioning~\cite{lucasius1994understanding} makes sense in this regard.
Detailedly, the set of values of parameter $\mu$ is partitioned into a number of smaller subsets so that
genetic searches can be conducted for each separate partition one after another. The partial solutions of these searches are
subsequently combined to obtain an estimated final solution.  
We need to note that there are no entirely-independent partitions in real-world problems; 
instead, the simulation states between the adjacent partitions are correlated in one way or another.
The correlation between two adjacent partitions should thus be correctly identified (visit the \texttt{sims} data in Sect.~\ref{hz:system-init}); otherwise, the estimated final solution may not be accurate.
The \textit{sliding window} (SW) technique~\cite{datar2002maintaining,babcock2002sliding} has been widely used in applications entailing the processing over data streams to perform data segmentation.
This technique makes automated analysis of a large stream of data durable in such a way as to discount past data items (already being analyzed) and only consider the relevant ones in the current window.
The sliding window is classified into two types: non-overlapping and overlapping~\cite{dehghani2019quantitative}.
The overlapping sliding window (OSW) is characterized by two parameters -- \texttt{opt\_window\_size} and \texttt{opt\_shift\_size}.
The former denotes the window size and the latter signifies the shift/overlapping size, which should be less than the
given window size. The non-overlapping sliding window (NOSW) can thus be seen as a special overlapping one, where the shift size is $0$.

This article aims to develop a compute-based implementation of framework/tool based on a novel calibration method/algorithm combining the \textit{single-objective} parallel EGA and OSW for the SEIR models extended with time/spatial-varying parameters.
Given the feasibility of running our calibration method within acceptable computation time, we implement it in 
Fortran programming language~\cite{GFortranDocu} that is extensively used in numerical and scientific computing since it can be highly optimized to run on high-performance computers, and in general the language is suited to producing parallel code (e.g., parallel EGA) where performance is important. Our calibration method also uses the OSW technique to divide the complex problem into overlapping sub-problems embracing different partitions. To increase the calibration accuracy, it incorporates \textit{directional information} that aids
in dynamically refining the ranges in which the values of parameter $\mu$ are allowed to vary and then guides the search space of the next partition in the direction of convergence, as the window slides forward.
We further propose a framework/tool attached with input/output modules in which the user can flexibly insert their data source, SW-related and GA-related parameters, and consume the ultimate solution containing a set of near-optimal values of the parameter.
In this way, the implementation of the method is treated as a black box, which is abstracted by our framework away from the user. The problem inherent in the OSW-based calibration implementation is that the solution at a certain point in time cannot be tuned again at a later time once it is determined, even if it is not satisfactory~\cite{babcock2002sliding}.
It can be addressed by imposing a post-tuning operation, i.e., subsequent improvement of the incorrect solution with a GA operation.
Therefore, our calibration framework supports two types of calibration events: automated calibration and \textit{one-off calibration}.
The automated calibration is supposed to adjust in accordance with directional knowledge and automates
over a given period of time without any empirical
interference from the user side. On the other hand, the one-off calibration enables post-tuning of the undesirable solution 
by performing a GA operation at given points in time.



In the next section, we describe an extended SEIR model used to test our proposed calibration algorithm. Sect.~\ref{hz:methods} describes the general methodology for the calibration algorithm. In Sect.~\ref{hz:framework} we then introduce the structure of the calibration framework used for the optimization and demonstrate the utility of the framework to calibrate the extended SEIR Model for representing the hospitalization dynamics related to COVID-19 disease in Germany. We show the results comparing the simulated dynamics and observed data in Sect.~\ref{hz:res}. Finally, we conclude in Sect.~\ref{hz:conclusion}.

\section{CoSMic model}
\label{hz:cosmic}
The model chosen for testing our calibration method is
an extended SEIR model, which is proposed in a recent research study~\cite{klusener2020forecasting}.
This extended model specifically takes sub-national/spatial and temporal variation in ICU-relevant disease
dynamics into account. It is also called an SEIR-based age-structured spatially-disaggregated
microsimulation model, which is hereafter abbreviated as \textit{CosMic}.
CosMic is designed and run for simulating and forecasting COVID-19-related ICU demand at the state level.   

A crucial technical challenge in applying it is its accurate calibration against observed data, e.g.,
the daily number of ICU patients with COVID-19, at the sub-regional/NUTS-2 regions,
as the history of the disease strongly affects predictions of future system behavior.
Specifically, the calibrated CoSMic represents a near-optimal estimation of the (floating-point) parameter \textit{R0\_effects}/$\mu$
that impacts the reproduction number $R_{0}$. The values of parameter $\mu$ vary and depend on miscellaneous factors --
seasonal change, location, pharmaceutical- and political-interventions, among others. I.e., 
the parameter $\mu$ varies as the time lapses and the region changes.
The rest of this paper uses the R0\_effects and $\mu$ interchangeably.
 
There are altogether $38$ NUTS-2 regions across Germany and initially the values of $\mu$ range from $0.1$ to $0.9$. 
The reported ICU case data set was extracted from the novel German intensive care register database established in March 2020 by the German Interdisciplinary Association for Intensive Care and Emergency Medicine (DIVI), since the real-life implementation is anticipated.


\section{Methods}
\label{hz:methods}
In our study, we applied the Genetic Algorithm (GA) to the calibration of our CoSMic model by using the sliding window technique. Therefore, before getting into the details of the calibration method, the concept and structure of the sliding window technique and GA are briefly introduced and their representation in the CoSMic model is illustrated as well.

\subsection{Overlapping sliding window}
\label{hz:osw}

We let notation 
\begin{equation} \label{hz:slidewindow}
U = <\overrightarrow{\mu_1},\overrightarrow{\mu_2},\overrightarrow{\mu_3},...,\overrightarrow{\mu_N}>
\end{equation}
be an ordered stream of \textit{regular time series} data~\cite{esling2012time}, which consists of data points over N time steps/weeks. Each data point is represented
as a vector used for storing a sequence of optimized values of our targeted parameter $\mu$ for different regions at a given time instant. These data points are not measured independently, but with \textit{one-way dependence}, meaning that the estimated value of data point $\overrightarrow{\mu_x}$ in CoSMic strongly depends on the model state expressed by its predecessor $\overrightarrow{\mu_{x-1}}$, not vice versa.
This is due to that the observed data, to which our model CoSMic is calibrated, is relevant to the COVID-19 patients admitted to the ICU. \textit{ICU-ill} is an intermediate state in the CoSMic model~\cite{klusener2020forecasting} and thus the number of current patients with ICU-ill state is estimated based on that of the earlier ones with other states (such as susceptible, exposed and infected).
Theoretically, optimizing $U$ at a single calibration execution is the simplest way to get a complete solution since we do not need to explicitly store any model state during calibration. However, it is unrealistically computationally expensive.
Therefore, the SW technique is applied to perform the problem partition~\cite{lucasius1994understanding}. Individual calibrations are then conducted for solving segmented problems one after another. Consequently, the solution of every single problem combines to determine the best search trajectory of the spatial-/time-varying parameter $\mu$.

\begin{figure}[h]
\sidecaption
 \includegraphics[width=.5\textwidth,height=.09\textheight]{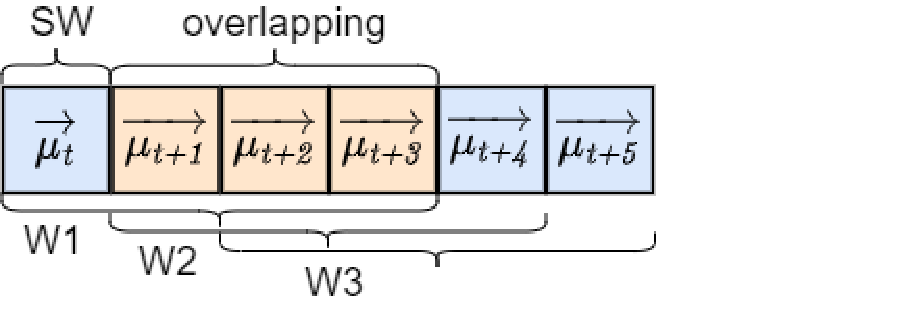}
\caption{The schematic of segmentation and overlapping sliding overlapping window}
\label{hz:sw_schematic}       
\end{figure}

The SW is proven to be efficacious in tackling optimization problems where the parameters of a model might vary over time~\cite{akhavizadegan2021time,ratnavale2022sliding}.
This technique considers only the active data items in the current window that nicely complies with the principle 
of one-way dependence.
With the same window size, segmenting using OSWs is almost twice longer than that using NOSWs.
Likewise, the estimation time on the data encompassed by OSWs is thus almost two times longer compared to the NOSWs.
We demonstrate a bias in favor of OSWs, notwithstanding such an increase in size and computation.
This is due to that OSW enables a more correct approximation by allowing for the application of additional knowledge to further refine some
of the active data items in the precedent window (see Sect.~\ref{hz:calibration-alg}).
Naturally, the two parameters -- \texttt{opt\_window\_size} and \texttt{opt\_shift\_size} are key factors
affecting the approximation performance (i.e., accuracy) of the proposed calibration method.
We tried varied values for them in this study, and the combination of $4$ and $1$ resulted in the best performance.
Fig.~\ref{hz:sw_schematic} exemplifies the application of the OSW-based segmentation to segment the calibration procedure from $t$-th week to $(t+5)$-th week.


\subsection{Genetic algorithm}
\label{hz:ga}
Genetic algorithms (GAs)~\cite{holland1992genetic} are population-based metaheuristics based on evolutionary processes of natural systems, which were invented by John Holland~\cite{sampson1976adaptation} in 1960s; hereafter they are extensively studied and developed by Goldberg~\cite{goldberg1989zen} and De Jong~\cite{de1975analysis}, among others.
Basically, GAs are used to find global near-optimal solutions to optimization and search problems by iteratively
applying evolution concepts including selection, mating, mutation and "survival of the fittest".
The calculation of fitness value is done by executing the defined fitness function.
A simply-implemented GA first initializes a random-based population comprising a set of individuals/\textit{chromosomes}, and then evaluates
the fitness for each individual. After that, it selects the parent having the strongest fitness, followed by crossover, and mutation operations to work out a new population. Finally, it calculates the fitness of each new individual and selects the next generation. This process is repeated until any of the predefined termination criteria is met. Hereafter the terms individual and chromosome are used interchangeably.

\subsubsection{Representation}
\label{hz:representation}

\begin{figure}[h]
\sidecaption
 \includegraphics[width=.75\textwidth,height=.08\textheight]{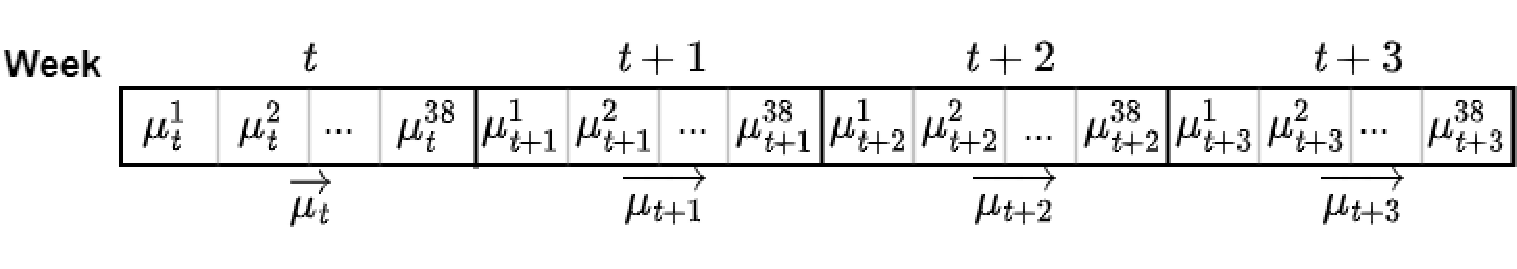}
\caption{Chromosome representation of a candidate solution within certain time window}
\label{hz:chromosome}       
\end{figure}

The first step in utilizing GAs for a particular problem is to design an appropriate representation of chromosome.
In this study, our targeted optimization problem (i.e., CoSMic) concerns the evaluation of parameter $\mu$ at the level of $38$ NUTS-2 regions over time. 
The best experimental values of \texttt{opt\_window\_size} and \texttt{opt\_shift\_size} are $4$ and $1$, respectively, and the $\mu$ is a real-valued continuous variable;
thus the chromosome is coded in the form of a vector of 38$\times$4 floating point numbers/\textit{genes}~\cite{golub1996implementation},
as shown in Fig.~\ref{hz:chromosome}.

The regions of interest are numbered in ascending order starting from $1$ up to $38$. Each gene $\mu_{t}^{s}$ alters the basic reproduction number $R_{0}$ and is calibrated so that our model results
for regions $s$ at COVID-19 week point $t$ can best reproduce the corresponding observed data. 
All the genes in the same week comprise a subset and then all these disjoint subsets $<\overrightarrow{\mu_{\textit{t}}},\overrightarrow{\mu_{\textit{t}+1}},\overrightarrow{\mu_{\textit{t}+2}},\overrightarrow{\mu_{\textit{t}+3}}>$ form the solution to the listed sub-problem constrained by the span of the current window. 
Initially, each gene takes value in a predefined interval $[0.1,0.9]$, which is problem-dependent and thus provided by the input parameter file.
The structure of chromosome representation remains unchanged as the sliding window moves; 
but we change to a new sub-problem embracing different optimized targets. Their values are subject to updated lower and upper bounds, which are
determined according to the model results at the previous week. 

Our representation scheme can be directly passed to the objective function (see Equation~\ref{hz:rmse}) without any intermediate encoding or decoding steps.
Further, the genetic operators (mutation and crossover) are accordingly defined to handle floating-point numbers 
(refer to Sect.~\ref{hz:structureofGA}).

\subsubsection{The objective function and fitness function}
\label{hz:obj} 

Aside from fixing a proper representation, the user must specify the objectives for initiating GA.
Our targeted CoSMic model produces a critical simulation output -- per-day ICU data, against which the model is calibrated.
The per-day ICU (hospitalization) data can intrinsically reflect the epidemic dynamics because of their reliability.
Thus a time slice mathematical representation of the CoSMic model is as follows:
\begin{equation} \label{hz:model}
CoSMic(<\overrightarrow{\mu_\textit{t}},\overrightarrow{\mu_{\textit{t}+1}},\overrightarrow{\mu_{\textit{t}+2}},\overrightarrow{\mu_{\textit{t}+3}}>) = \overrightarrow{ICU^{sim}},
\end{equation}
where the vector $<\overrightarrow{\mu_\textit{t}},\overrightarrow{\mu_{\textit{t}+1}},\overrightarrow{\mu_{\textit{t}+2}},\overrightarrow{\mu_{\textit{t}+3}}>$ is the parameters to be calibrated, and the vector $\overrightarrow{ICU^{sim}}$ is the
simulated ICU cases.



Essentially, the parameter calibration of the above time slice CoSMic model means adjusting the values of parameters that can make the ICU data match the observed one in the course of \textit{t}-th to (\textit{t}+3)-th week closely.
There are a set of closeness indicators used to evaluate the goodness-of-fit of the calibration but we have
used only root mean square error (RMSE~\cite{chai2014root}) for our study.
A smaller value of RMSE indicates a set of parameters introducing a closer approximation to the observed data.
Therefore, we formulate our objective function as follows:
\begin{equation} \label{hz:rmse}
Obj = \sqrt{\sum\limits_{i=1}^{m}\sum\limits_{j=1}^{n}(ICU_{i,j}^{obs} - ICU_{i,j}^{sim})^2/(m*n)},
\end{equation}
where in this study "m=28" indicates the 28-day window and "n=38" represents the number of regions.
The $ICU_{i,j}^{obs}$ and $ICU_{i,j}^{sim}$ are the observed and simulated ICU cases on the $j$th region at $i$-th day of the
predefined time window, respectively.
In the search process of GA, the individuals with higher fitness values (i.e., smaller RMSE) are more likely to survive.
Therefore, the fitness function is further created by simply applying a negative to the above objective function:
\begin{equation} \label{hz:fitness}
Fitness = -Obj
\end{equation}


\subsubsection{Implementation/Structure of proposed GA heuristic}
\label{hz:structureofGA}
We apply Elite GA (EGA) so as not to lose the few best-found solutions. It improves "simple" GA in a way
that passes elite individuals (i.e., individuals with the largest fitness values) on to the next generation without
crossover and mutation operations, unless new individuals with larger fitness are generated and selected
as the new elite individuals (the maximum number of elite individuals applied was $5$ in this study).
The main steps involved in our implementation of EGA can be summarized as follows:
\begin{enumerate}[label=\arabic*.]
\item \textbf{Initialization}: construct N candidate solutions to form an initial population. The initial population constitutes base solutions for successive generations. Each candidate solution (chromosome) is generated randomly and the values of its genes are chosen by obeying a uniform distribution between the given lower and upper bounds.
\item \textbf{Fitness evaluation}: calculate the fitness value for each individual in the population according to the fitness function
expressed in Equation~\ref{hz:fitness}. Then all the individuals are sorted in ascending order of the fitness values. 
The individuals with higher fitness values are thus regarded as fitter candidate solutions.
\item \textbf{Elitism}: select individuals with the largest fitness values as the elite individuals, which will be
straightforwardly passed on to the next generation without crossover and mutation operations.
\item The following steps are repeated until there are N individuals in the population of the next generation.
\begin{enumerate}[label*=\arabic*.]
\item \textbf{Selection}: select good candidate solutions as parents to reproduce the next generation of the population. There are different
strategies for selection but here we have used only roulette wheel selection~\cite{de1975analysis}. The idea of roulette wheel selection is to assign a probability value to each chromosome based on its fitness value, and then select two chromosomes as the parents according to the probability distribution. The chromosomes having the higher fitness value have a better chance to be selected; thus 
a chromosome could repeatedly be selected.
\item \textbf{Crossover/Recombination}: in order to generate offspring solutions, a crossover operator is applied to the selected parents with the given crossover probability (80\% for our experiment). The proposed algorithm uses a whole arithmetic crossover strategy, which is useful with floating-point solutions. 
It takes a percentage of each parent gene and adds them to produce new solutions (two children). The value of the percentage is randomly generated.
The newly produced children are inserted into the population of the next generation.
\item \textbf{Mutation}: mutation enhances diversity at a smaller probability value (10\% for our experiment). 
We have used a simple procedure where we randomly choose one gene from a chromosome and change its value randomly.
\end{enumerate}
\item Repeat steps 2-4 until termination criteria are satisfied. In our study, the algorithm stops when the maximum number-of-generations or the convergence-of-population is achieved. Finally, the best-found solution is presented.
\end{enumerate}
The Fortran version of our calibration framework has been developed based on the above-described EGA.
The values of algorithm-related parameters need to be specified beforehand (visit the input module in Fig.~\ref{hz:syntheticwf}) to define a specific heuristic search.

\section{Proposed framework}
\label{hz:framework}
In this section, we introduce the new framework\footnote{https://github.com/hpcralf/CoSMic} for calibrating SEIR models extended with the characteristic of time-/spatial-variation (like CoSMic). This framework encapsulates our calibration algorithmic details of integrating the GAs and OSW as core component. Besides, our framework provides input and output components, where the user group can flexibly insert the configurable parameters and consume the optimized solution.

\subsection{High-level structure}
\label{hz:highlevel}

\begin{figure}[ht]
\sidecaption
\includegraphics[width=.6\textwidth,height=.35\textheight]{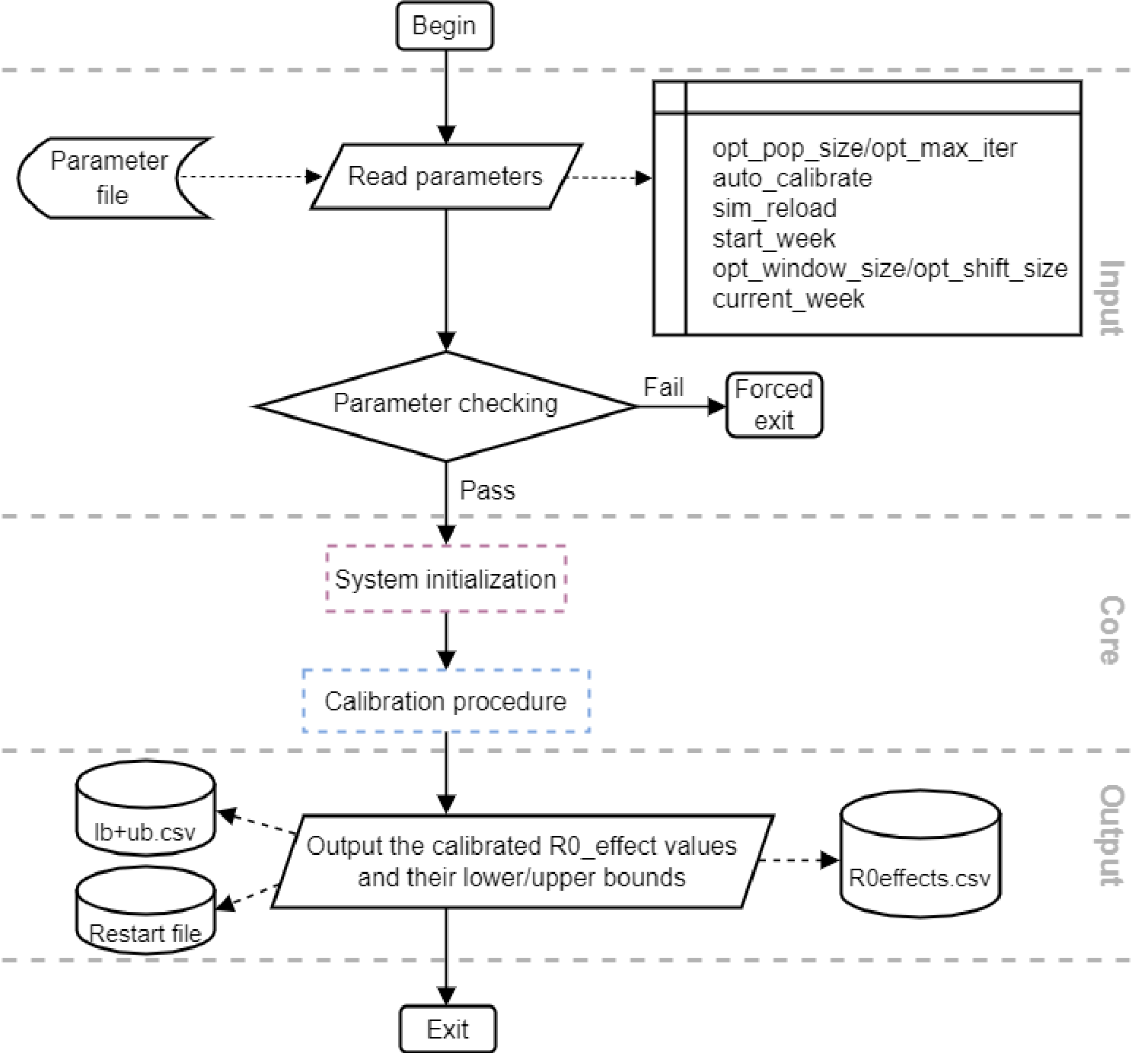}
\caption{The synthetic workflow of the proposed calibration framework}
\label{hz:syntheticwf}       
\end{figure}

Fig.~\ref{hz:syntheticwf} demonstrates a high-level (abstract-level) overview of our calibration framework structure
mainly consisting of three sub-components:
\begin{enumerate*}[label=\emph{\alph*)}]
  \item input,
  \item core, and
  \item output
\end{enumerate*}.
The input component reads all the needed parameters from a \texttt{static\_parameter.dat} file and prepares them for initializing the framework. The input validation is called for the detection of unauthorized input before it is passed on to the following components.
It is applied on the semantic level -- enforces the correctness of the parameters' values in the context of CosMic use case.
The input parameters fall into three different categories:
\begin{enumerate*}[label=\emph{\alph*)}]
  \item automation-relevant,
  \item system, and
  \item GA-relevant/control
\end{enumerate*}.
In detail, the automation-relevant parameters target the data segmentation and then the automation procedure that is performed by the SW technique. They can be enumerated as follows:
\begin{enumerate*}[label=\emph{\arabic*)}]
  \item \texttt{auto\_calibrate}: a logical value indicating an automated (TURE) or a one-off calibration procedure (FALSE),
  \item \texttt{opt\_window\_size}: window size,
  \item \texttt{opt\_shift\_size}: shift factor,
  \item \texttt{current\_week}: the window stops sliding when \texttt{current\_week} reaches, and
  \item \texttt{start\_week}: the start point of the first window
\end{enumerate*}.
The system parameter here is \texttt{sim\_reload}, which is a logical value indicating a restart file reload (TURE) or not (FALSE).
The timescales of \texttt{opt\_window\_size} and \texttt{opt\_shift\_size} are expressed in weeks.
In the beginning, the values of parameter \texttt{start\_week} and \texttt{sim\_reload} are respectively assigned $1$ and FALSE, which indicate
the initial state of our automatic calibration event.
The GA-relevant parameters are featured by the following control and range-related indexes:
\begin{enumerate*}[label=\emph{\arabic*)}]
  \item \texttt{opt\_pop\_size}: population size,
  \item \texttt{opt\_max\_size}: number of generations,
  \item \texttt{opt\_lb}: the lower bound of the parameter, and
  \item \texttt{opt\_ub}: the upper bound of the parameter
\end{enumerate*}.

System initialization and calibration procedure comprise our core component.
They contain a level of steps that combine to perform a relatively complicated routine 
and thus are unfolded separately in Sections~\ref{hz:system-init} and \ref{hz:calibration-alg}.
In the end, the best solution (optimized values of parameter $\mu$ from \texttt{start\_week} to \texttt{current\_week})
is written to an output file \texttt{R0effects.csv}, which is ready to be accessed in the further.
The directional information indicating the upper and lower bounds of the parameter $\mu$ for the next round calibration is
recorded into an output file \texttt{lb+ub.csv}.
Besides, the relevant restart file, from which the next calibration event starts, is outputted.

\subsection{System initialization}
\label{hz:system-init}
\begin{figure}[h]
\sidecaption
 \includegraphics[width=.6\textwidth,height=.47\textheight]{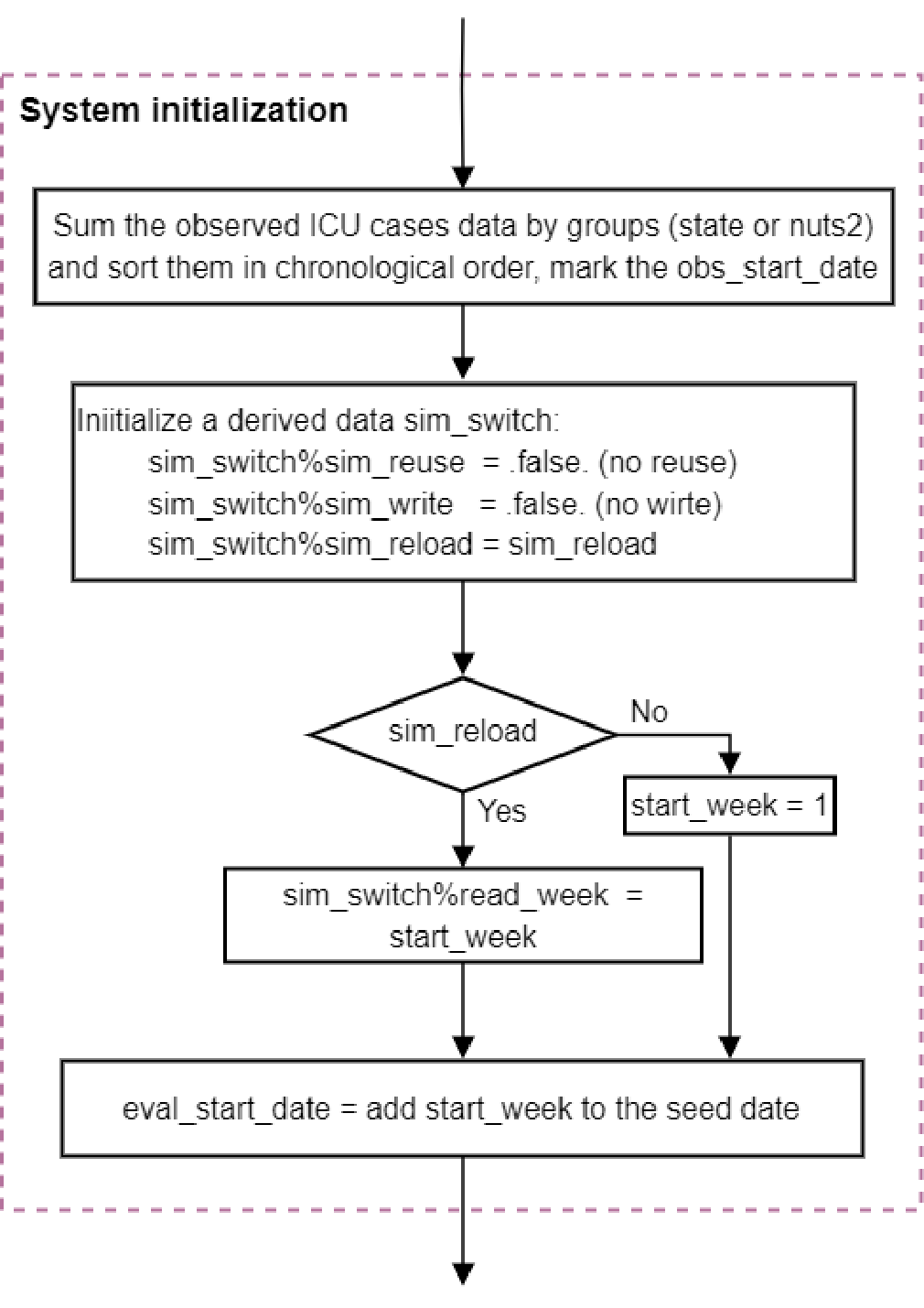}
\caption{The workflow of system initialization component}
\label{hz:systeminiwf}       
\end{figure}

Fig.~\ref{hz:systeminiwf} dissects the system initialization component with a sequence of initializing actions.
The observed/referred ICU cases are summed by regions (either state or NUTS-2) and then organized in chronological order.
And then \texttt{sim\_switch} of the derived (\texttt{struct}) data type is initialized before it is passed and analyzed in the simulation module.
Its member \texttt{sim\_reuse} is initialized with FALSE implying that the involved simulation starts from scratch or
reloads a restart file. 
Once the window starts to slide forward, the \texttt{sim\_reuse} is changed to TRUE meaning that
the simulation receives the passed \texttt{sims} data, which is understood as the previous model state and perceived as the new starting point. 
The other members will be introduced on demand in the future.
When the simulation entails a reload of a restart file (with \texttt{sim\_reload} as true), the specific \texttt{start\_week} is provided to indicate from which week
this simulation starts. Accordingly, the starting evaluated date represented by this simulation should be adjusted by adding \texttt{start\_week} to the seed date (March 9, 2020), i.e., the first day in week $1$ of the German COVID-19 epidemic.

\subsection{Calibration algorithm}
\label{hz:calibration-alg}
\begin{figure}[h]
\sidecaption
 \includegraphics[width=.64\textwidth,height=.37\textheight]{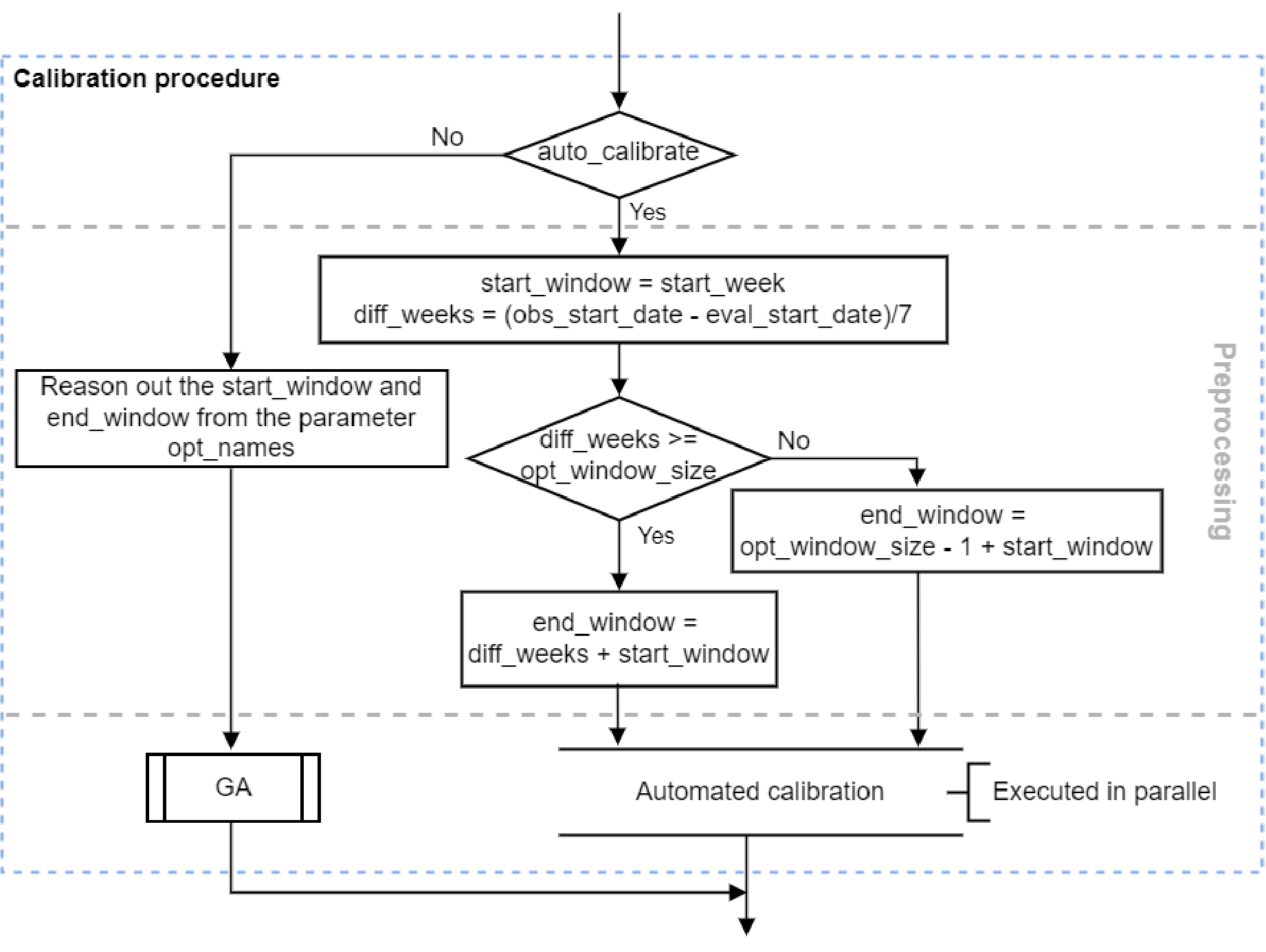}
\caption{The workflow of calibration procedure component}
\label{hz:calibrationprocedurewf}       
\end{figure}

The workflow of the calibration procedure is shown in Fig.~\ref{hz:calibrationprocedurewf}, from which we can obviously observe that
the calibration procedure goes in two distinct directions in terms of \texttt{auto\_calibrate}. 
The disabled \texttt{auto\_calibrate} indicates a one-off calibration which is more like a static post-tune of
the unsatisfied values by applying a GA.
The \texttt{start\_window} and \texttt{end\_window} are directly parsed from the parameter \texttt{opt\_names}. 
On the contrary, in the course of automated calibration,
the initial \texttt{start\_window} and \texttt{end\_window} are determined to include the observed starting date, and thus
the initial window size is not necessarily identical to the predefined \texttt{opt\_window\_size}.
The above sequence of steps is reduced to preprocessing.
The parallel routines -- GA and automated calibration -- come immediately after the preprocessing.
Next, we move to the details of automated calibration, on which we will focus in this Section.

\begin{algorithm}
\caption{Pseudo-Fortran code for solving automated calibration with OSW and GA for certain NUTS-2 region}\label{hz:calibrationcode}
  \begin{algorithmic}[1]
   \Function{Automated\_cali}{\textsl{R0\_effects, start\_window, current\_window, opt\_window\_size, opt\_shift\_factor, sims, Obs\_ICU, opt\_lb, opt\_ub}}
   \While{\textsl{start\_window $\leq$ current\_window}}
   \State \textsl{Generate initial population $\overrightarrow{P}$}
   \State \textsl{end\_window $\leftarrow$ start\_window+opt\_window\_size-1}
   \If{\textsl{end\_window $>$ current\_window}}
   \State \textsl{end\_window $\leftarrow$ current\_window}
   \EndIf
   \Repeat
   \State \textsl{R0\_effects(start\_window:end\_window,:) $\leftarrow$ $\overrightarrow{P}$}
   \State $\overrightarrow{Fitness}$ $\leftarrow$ \Call{Fitness\_cal}{\textsl{R0\_effects, start\_window, end\_window, sims, sim\_switch, Obs\_ICU}}
   \State \textsl{Selection enhanced with elitism, crossover, mutation in terms of $\overrightarrow{Fitness}$}
   \State \textsl{Update $\overrightarrow{P}$}
   \Until{\textsl{Stopping criteria reached}}
   \State \textsl{R0\_effects(start\_window:end\_window,0) $\leftarrow$ Best chromosome in $\overrightarrow{P}$}
   \State \Call{Adjust\_lbub}{\textsl{R0\_effects(start\_window:end\_window,0), sims, sim\_switch, Obs\_ICU, opt\_lb, opt\_ub}}
   \Comment{Adjust opt\_lb and opt\_ub in terms of the fitness value}
   \State \Call{CoSMic}{\textsl{R0\_effects(start\_window:end\_window-(opt\_window\_size-opt\_shift\_size),0), sims, sim\_switch}}
   \Comment{Record the sims data at week of end\_window-(opt\_window\_size-opt\_shift\_size)}
   \State \textsl{sim\_backup $\leftarrow$ sims}
   \State \textsl{sim\_switch\%sim\_reuse $\leftarrow$ .TRUE}
   \State \textsl{sim\_switch\%sim\_reload $\leftarrow$ .FALSE}
   \State \textsl{start\_window $\leftarrow$ start\_window+opt\_shift\_size}
   \State \textsl{end\_window $\leftarrow$ end\_window+opt\_shift\_size}
   \If{\textsl{end\_window $>$ current\_window}}
   \State \textsl{end\_window $\leftarrow$ current\_window}
   \Comment{Slide the window by opt\_shift\_size}
   \EndIf
   \EndWhile
   \State \textsl{return R0\_effects(:,0)}
   \State \textsl{Record the updated opt\_lb and opt\_ub}
   \State \textsl{Write the recorded sims data to a restart file}
   \EndFunction
   \Function{Fitness\_cal}{\textsl{R0\_effects, start\_window, end\_window, sims, sim\_switch, Obs\_ICU}}
   \State $\overrightarrow{Eva\_ICU}$ $\leftarrow$ \Call{CoSMic}{\textsl{R0\_effects(start\_window:end\_window,:), sims, sim\_switch}}
   \State $\overrightarrow{Fitness}$ $\leftarrow$ \textsl{RMSE($\overrightarrow{Eva\_ICU}$, Obs\_ICU)}
   \State \textsl{return $\overrightarrow{Fitness}$}
   \EndFunction
   \Procedure{Adjust\_lbub}{\textsl{R0\_effects, start\_window, end\_window, sims, sim\_switch, Obs\_ICU, opt\_lb, opt\_ub}}
   \State \textsl{Eva\_ICU} $\leftarrow$ \Call{CoSMic}{\textsl{R0\_effects(start\_window:end\_window,0), sims, sim\_switch}}
   \State \textsl{(diff,correct\_coeff)} $\leftarrow$ \Call{Adjust\_strategy}{\textsl{Eva\_ICU, Obs\_ICU}}\Comment{diff $\leftarrow$ Eval\_ICU(last day of end\_window)-Obs\_ICU(last day of end\_window)}
   \If{\textsl{diff .lt. 0}}\Comment{Lower than the observed data}
   \State \textsl{opt\_lb $\leftarrow$ avg\_R0}\Comment{avg\_R0 $\leftarrow$ mean(R0\_effects(start\_window:end\_window,0))}
   \State \textsl{opt\_ub $\leftarrow$ min(avg\_R0+correct\_coeff,0.9)}
   \ElsIf{\textsl{diff .gt. 0}}\Comment{Higher than the observed data}
   \State \textsl{opt\_lb $\leftarrow$ max(avg\_R0-correct\_coeff,0.1)}
   \State \textsl{opt\_ub $\leftarrow$ avg\_R0}
   \Else
   \State \textsl{opt\_lb $\leftarrow$ max(avg\_R0-correct\_coeff,0.1)}
   \State \textsl{opt\_ub $\leftarrow$ min(avg\_R0+correct\_coeff,0.9)}
   \EndIf
   \EndProcedure
  \end{algorithmic}
 \end{algorithm}
Algorithm~\ref{hz:calibrationcode} explains in detail the automated calibration procedure (for a given NUTS-2 region), where there is a two-tier nested loop.
In Line $10$, the fitness values for individuals can be evaluated independently in parallel. In our experiment, we use
the master-slave parallelism for the EGA~\cite{cantu1998survey}.
The inner \texttt{repeat-until} loop is entailed by the GA subroutine;
the outer \texttt{while} one accomplishes the automation procedure by moving forward the sliding window, until the predefined \texttt{current\_window} reaches.
The automation firstly goes through a GA subroutine, which starts from an initial population and then iteratively 
optimizes the \texttt{R0\_effects} from week \texttt{start\_window} to \texttt{end\_window} until the stopping criteria (see Line $13$) are satisfied.
In this GA subroutine, we evaluate the fitness value (as shown in Lines $30$-$34$) by calculating the root-mean-square error (RMSE), which represents the differences between the ICU cases estimated by CoSMic simulation and those observed between \texttt{start\_window} and \texttt{end\_window}.
This is followed by the sequence of selection enhanced with elitism, crossover and mutation. 
The selection is conducted in the principle of working out a new population from (fitter) individuals with higher fitness values.

After escaping from the GA loop, we call the simulation twice with the so far best-found solution (i.e., chromosome with the highest fitness value in $\overrightarrow{P}$) to prepare for the next calibration period. The preparation is twofold:
\begin{enumerate*}[label=\emph{\arabic*)}]
  \item One is to dynamically adjust the lower and upper bounds of the \texttt{R0\_effects} value (refer to the procedure \texttt{Adjust\_lbub} represented by Lines $35$-$48$) based on directional information mentioned in Sect.~\ref{hz:intro}. The adjustment strategy is thus
to some extent problem-specific but in principle should enable self-heuristic search space as the window slides.
In our case, the directional information combines the current solution (see Line $39$) and the 
difference between observed and evaluated ICU cases at the most recent point in time of the current window (see Line $37$).
The former can prevent the estimated $\mu$ from fluctuating strongly over adjacent periods and
yields good performance with low RMSE. The latter is justified by the approximately three-week delay between exposure 
to  SARS-CoV-2 and hospitalization due to COVID-19.
  \item The other is to memorize the \texttt{sims} data (i.e., simulation status) on the first day of a certain week from which the next calibration period starts (see Line $16$).
Next, we activate the \texttt{sim\_reuse}, slide the window by \texttt{opt\_shift\_size} and move to the next period if the 
start point of the current window does not exceed the \texttt{current\_week}.
In this regard, the called CoSMic simulation will use the passed \texttt{sims} data as its initial state after the first calibration period.
After escaping from the outer loop, the recorded \texttt{sims} data will be permanently written to a restart file which can be reloaded during the subsequent calibration procedures. Such strong \texttt{sims} data dependence will cause error propagation temporally. The occurrence of dramatic
drops/rises (peak/nadir) during the observed ICU data series could further deteriorate the error propagation. To alleviate it, a smoothing mechanism can be integrated to retouch the already-calibrated parameter values.
\end{enumerate*}

\section{Results}
\label{hz:res}

Our use case -- CoSMic model -- considered week-by-week variations in our time-varying parameter $\mu$, the first task
is thus to determine the values of two critical OSW-related parameters -- \texttt{opt\_shift\_size} and \texttt{opt\_window\_size}, which define how a series of optimized parameters should be split and then correlated. 
The calibration results may deviate from the observed data when the assumed \texttt{opt\_shift\_size} is longer than the timescale of actual variation.
We can recall that the best-combined values of \texttt{opt\_window\_size} and \texttt{opt\_shift\_size} are $4$ and $1$, by revisiting the Sect.~\ref{hz:osw}.
Table~\ref{hz:caliparameters} demonstrates the production setup (i.e., values of critical parameters) for the calibration procedure across
the NUTS-2 regions from March 9th 2020 (i.e., COVID-19 week $1$) to August 2nd 2022 (i.e., COVID-19 week $125$).
The sample size we choose is the full German population. 
The CoSMic model is stochastic and executed 40 times for each window (time period)
and then the estimated ICU cases are averaged across the multiple outputs in order to reduce variance in parameter estimation
and obtain a stable trend for it.

\begin{table}[h]
\caption{Calibration parameters and values}
\label{hz:caliparameters}       
\begin{tabular}{p{5cm}p{6cm}}
\hline\noalign{\smallskip}
Simulation parameters & Values \\

\noalign{\smallskip}\svhline\noalign{\smallskip}
Sample size & 83237124\\
Iteration numbers & 40\\
Calibrated weeks & COVID-19 week1-week125\\
Calibrated regions & All 38 German NUTS-2 regions\\
\noalign{\smallskip}\hline\noalign{\smallskip}
\end{tabular}
\end{table}

Our framework allows real-time calibration; thus we aim to obtain decent solutions within reasonable execution time
by assuming a limited number of iterations and population size. 
Table~\ref{hz:gaparameters} provides a layout of the GA user-defined configuration file used for calibrating the model parameters.
We can observe that the generation number is set as 10, since the fitness value level off as the iteration goes.
The population size is 80, which is large enough to guarantee diversity.
A crossover probability of $0.9$ and a mutation probability of $0.1$ are used in this simulation.
The upper and lower limit of the parameter $\mu$ for CoSMic model is read from the input files. 
Each chromosome is represented as a parameter array that comprises $4\times38$ (i.e., $152$) floating point numbers in their own range, given that there are altogether $38$ NUTS-2 regions across Germany. 


\begin{table}[h]
\caption{GA parameters and values}
\label{hz:gaparameters}       
%
%
\begin{tabular}{p{5cm}p{6cm}}
\hline\noalign{\smallskip}
GA parameters & Values \\
\noalign{\smallskip}\svhline\noalign{\smallskip}
\textbf{Population size} & 80\\
\textbf{Generation number} & 10\\
Mutation probability & 0.1\\
Crossover probability & 0.8 \\
Mutation method & Uniform Random\\
Selection type & Linear scaling\\
Crossover type & Logical arithmetic\\
Elitism & 5\\
Chromosome size & 152 floating point numbers\\
\noalign{\smallskip}\hline\noalign{\smallskip}
\end{tabular}
\end{table}

\begin{figure*}
\centering
\includegraphics[width=.8\textwidth,height=.25\textheight]{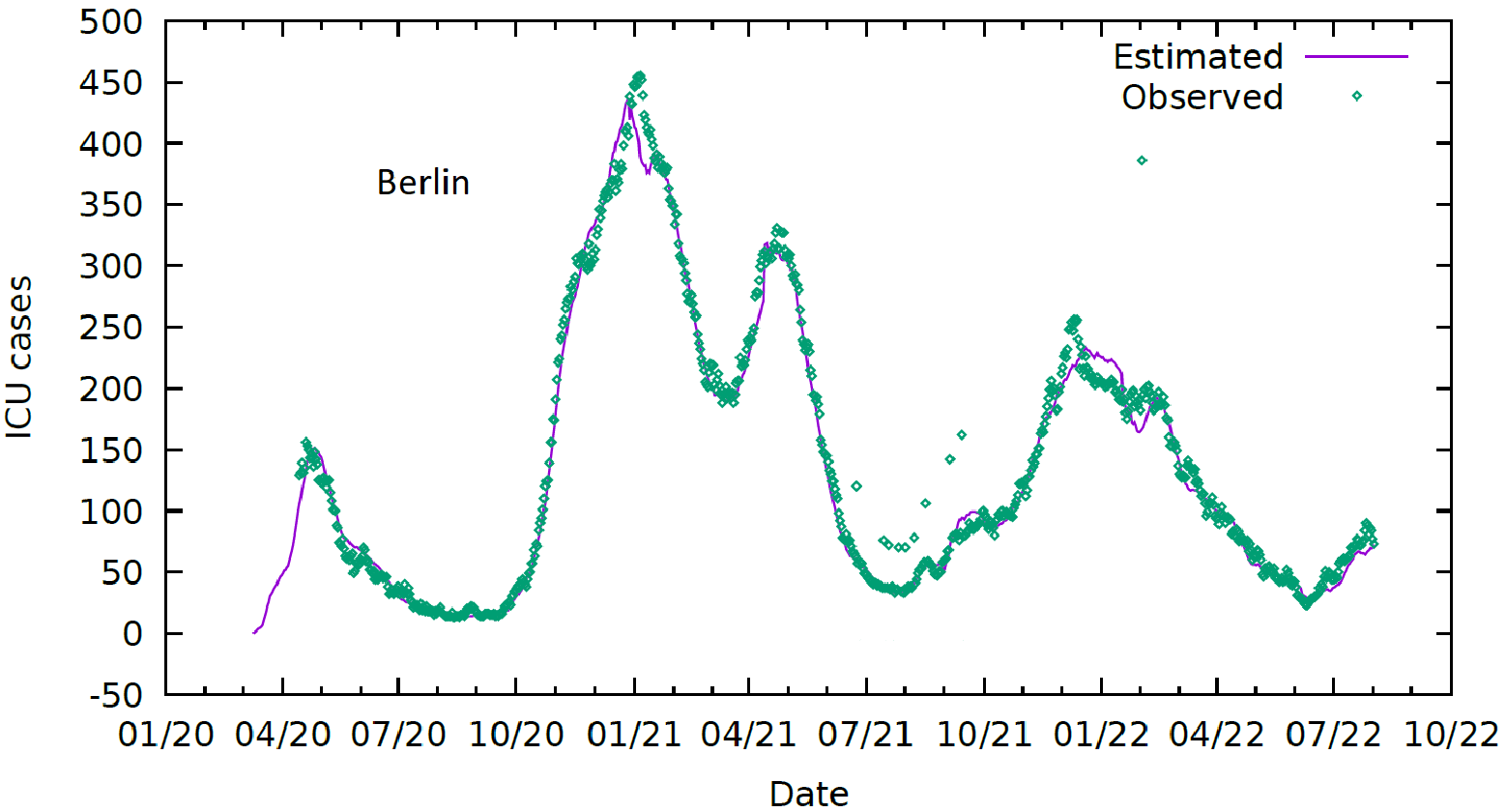}\quad
\includegraphics[width=.8\textwidth,height=.25\textheight]{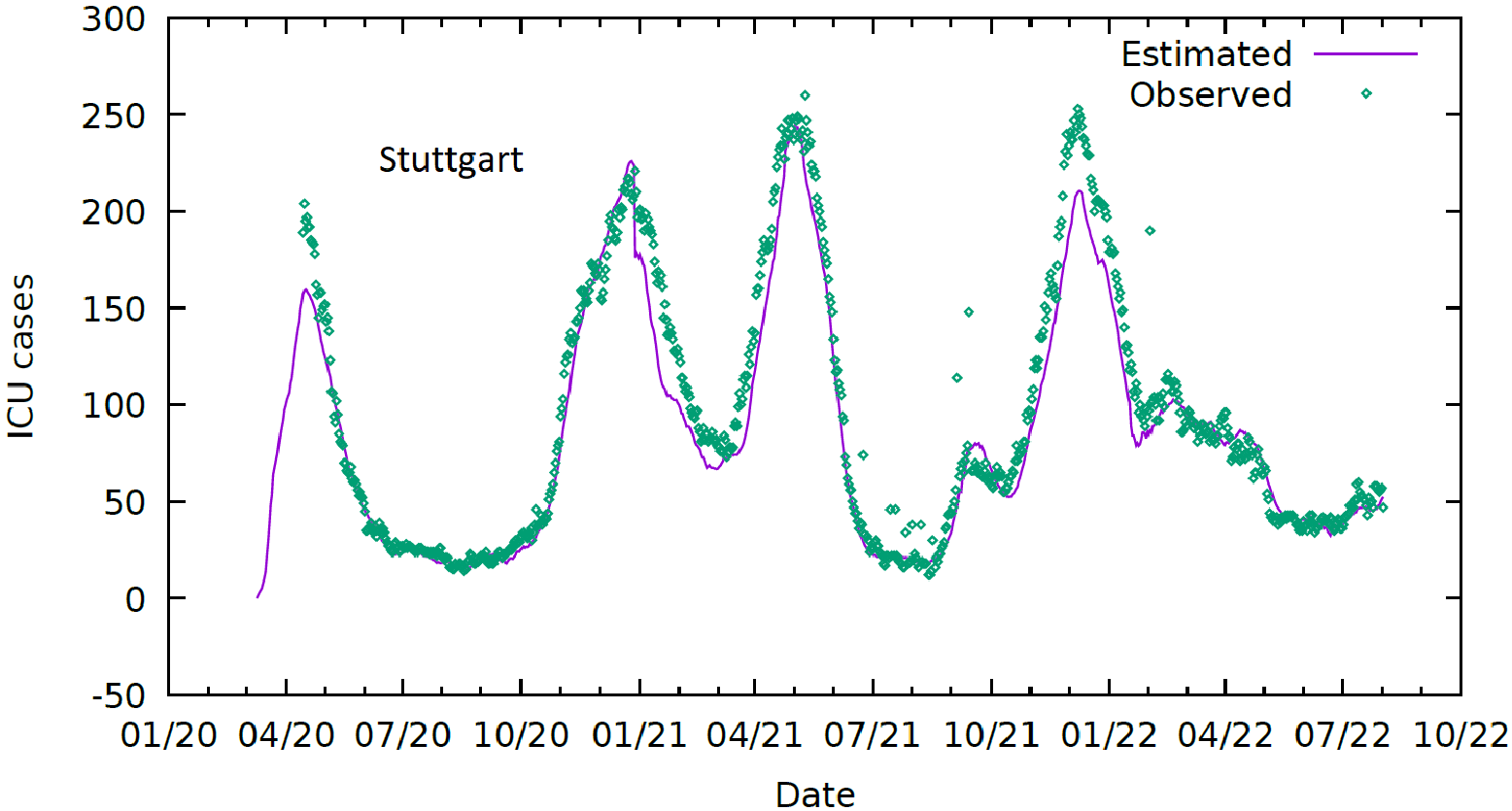}\par\medskip
\includegraphics[width=.8\textwidth,height=.25\textheight]{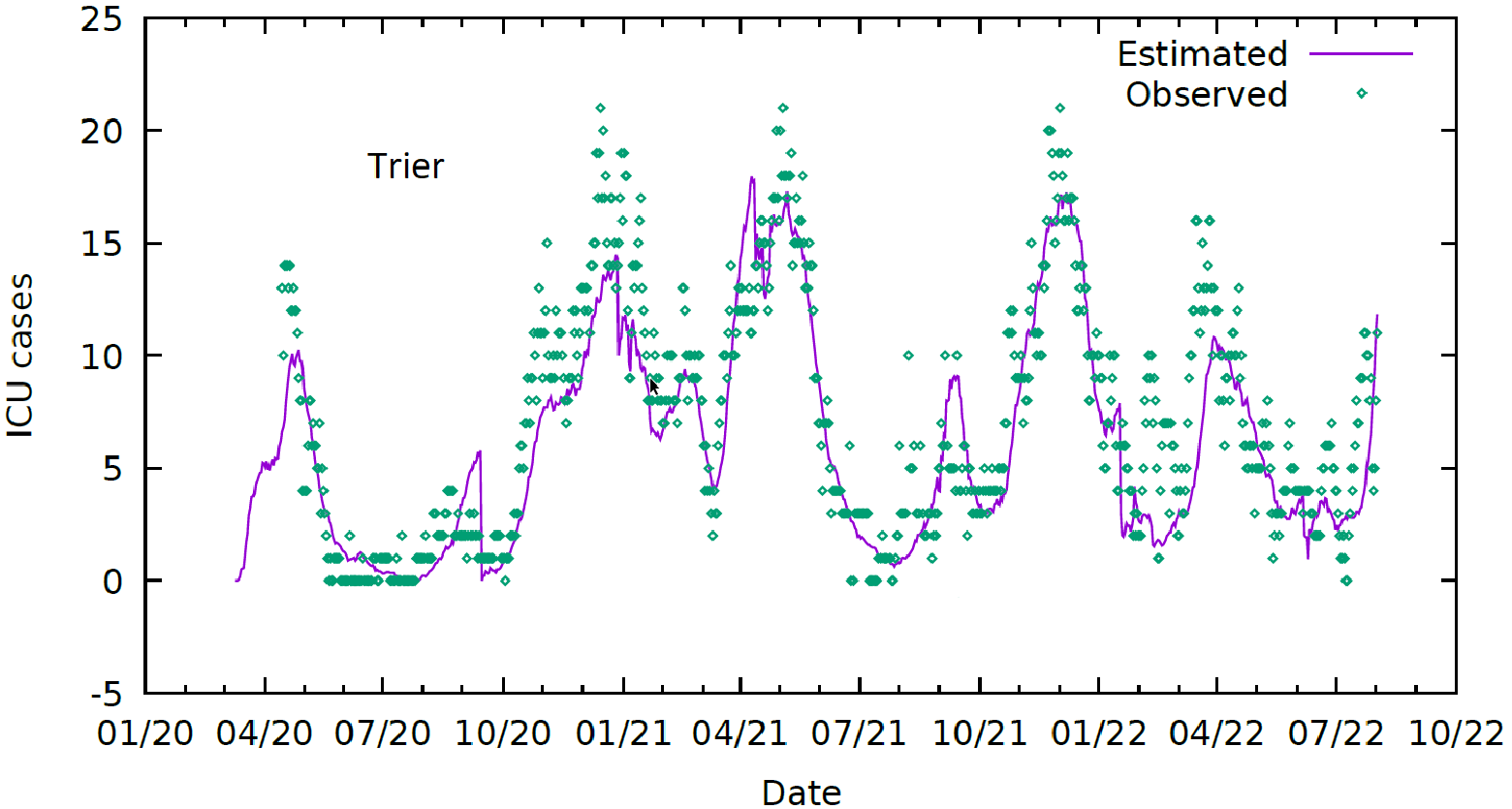}
\caption{Model-fitting of weekly transmission coefficient $\mu$ and daily hospitalizations in regions of Berlin, Stuttgart and Trier.}
\label{hz:comparisoncurve}
\end{figure*}

The CoSMic model is not a hypothetical example, where the true optimum set of parameter values was known by assumption.
We cannot use this to examine whether our calibration method is capable of finding the optimum.
Therefore the accuracy of our calibration method was then studied by checking with the data fitting, i.e., 
the difference (RMSE) between the model-estimated ICU dynamics and the observed one.
Next, we will show how our framework performs in calibration accuracy and speed under the production setup (see Table~\ref{hz:caliparameters}) and the given GA-related parameters (see Table~\ref{hz:gaparameters}).
As we mentioned in Sect.~\ref{hz:representation}, initially the GA search range of $\mu$ is set to [0.1,0.9]. 
Among all the $38$ regions in Germany, Berlin, Stuttgart and Trier are typical of regions with low, medium and high density, respectively.
Fig.~\ref{hz:comparisoncurve} plots the comparison curves in the three regions for our single objective (i.e., to minimize the average RMSE between the model-estimated ICU dynamics and the observed data).
A high match between the estimated average and observed ICU cases can be obviously observed in Berlin and Stuttgart regions and the important overall patterns (e.g., uphills, downhills, etc.) can consistently be captured in a satisfactory way. The average RMSE is in the order of $12$ over all the NUTS-2 regions and the past two years, which is low given that most values of our dataset are larger than 100, and indicates a good calibration accuracy.

We ran the above calibration procedure (refer to Table~\ref{hz:caliparameters}) on the NEC cluster (aka. Vulcan) located at HLRS. Vulcan consists of multiple nodes of different types.
The compute node type we used is Intel Haswell E5-2660v3. Each of the Haswell compute node has 20 cores running at \SI{2.6}{\giga\hertz} with \SI{128}{\giga\byte} DDR3 main memory. The applied GNU compiler version was $11.2.0$ and the version of OpenMPI/$4.1.2$ was run.
In this experiment, 80 parallel nodes are requested and a single MPI process is launched per node. Each MPI process further spawns $20$ OpenMP threads that concurrently execute the $40$ CoSMic model runs for each generation. 
Owing to the parallelized implementation, this calibration procedure took approximately 1.5 days to complete.

\section{Conclusion}
\label{hz:conclusion}
We explore a calibration method to fit an extended SEIR model of COVID-19 to location-specific ICU data for estimating spatial-/time-varying parameters. 
This calibration method combines the OSW technique and EGA to achieve satisfactory estimation accuracy within a reasonable time.
On the one hand, the applied OSW technique is used to segment a given calibration procedure into multiple ones
over shorter time slices; thus it enables an automated process of parameter estimation by moving forward the window.
On the other hand, the EGA is an improved global optimization method in high-dimensional spaces and is applied to each segmented calibration.
Naturally, it is simple to be parallelized in code and becomes highly effective by simultaneously calculating the fitness values of multiple individuals when executed in parallel. 
Besides, we use a high-performance computing environment to foster many GA heuristic search processes in parallel, allowing us to effectively explore a broad parameter space in a time-efficient manner.
Further, the implementation details of this algorithm are wrapped in our calibration framework and thus hidden from the user.
Our framework offers the input/output components, where the user can insert the calibration and EGA settings of interest
and consume the ultimate solutions at his/her convenience as well.
Overall, our calibration framework shows promise for estimating spatial-/time-varying parameters and working
with the models with different settings.
In the future, we envision that the works include
\begin{enumerate*}[label=\emph{\alph*)}]
\item strengthen the validation of our calibration algorithm by comparing it
with the existing calibration algorithms,
\item accelerate the optimization speed, and
\item improve the optimization accuracy
\end{enumerate*}.
\begin{acknowledgement}
We would like to acknowledge the funding provided by the BMBF \& MWK Baden W\"urttemberg through the work package "generic scenarios and prototypical implementations" (AP2) in the project CIRCE (under Grant no. 16HPC062).
\end{acknowledgement}

\bibliographystyle{spmpsci}
\bibliography{./ref.bib}

\begin{thebibliography}{10}
\providecommand{\url}[1]{{#1}}
\providecommand{\urlprefix}{URL }
\expandafter\ifx\csname urlstyle\endcsname\relax
  \providecommand{\doi}[1]{DOI~\discretionary{}{}{}#1}\else
  \providecommand{\doi}{DOI~\discretionary{}{}{}\begingroup
  \urlstyle{rm}\Url}\fi

\bibitem{akhavizadegan2021time}
Akhavizadegan, F., Ansarifar, J., Wang, L., Huber, I., Archontoulis, S.V.: A
  time-dependent parameter estimation framework for crop modeling.
\newblock Scientific reports \textbf{11}(1), 1--15 (2021)

\bibitem{babcock2002sliding}
Babcock, B., Datar, M., Motwani, R., O'Callaghan, L.: Sliding window
  computations over data streams.
\newblock Tech. rep., Stanford InfoLab (2002)

\bibitem{barnhart2017moesha}
Barnhart, B.L., Sawicz, K.A., Ficklin, D.L., Whittaker, G.W.: Moesha: A genetic
  algorithm for automatic calibration and estimation of parameter uncertainty
  and sensitivity of hydrologic models.
\newblock Transactions of the ASABE \textbf{60}(4), 1259--1269 (2017)

\bibitem{cantu1998survey}
Cant{\'u}-Paz, E., et~al.: A survey of parallel genetic algorithms.
\newblock Calculateurs paralleles, reseaux et systems repartis \textbf{10}(2),
  141--171 (1998)

\bibitem{chai2014root}
Chai, T., Draxler, R.R.: Root mean square error (rmse) or mean absolute error
  (mae)?--arguments against avoiding rmse in the literature.
\newblock Geoscientific model development \textbf{7}(3), 1247--1250 (2014)

\bibitem{datar2002maintaining}
Datar, M., Gionis, A., Indyk, P., Motwani, R.: Maintaining stream statistics
  over sliding windows.
\newblock SIAM journal on computing \textbf{31}(6), 1794--1813 (2002)

\bibitem{de1975analysis}
De~Jong, K.A.: An analysis of the behavior of a class of genetic adaptive
  systems.
\newblock University of Michigan (1975)

\bibitem{dehghani2019quantitative}
Dehghani, A., Sarbishei, O., Glatard, T., Shihab, E.: A quantitative comparison
  of overlapping and non-overlapping sliding windows for human activity
  recognition using inertial sensors.
\newblock Sensors \textbf{19}(22), 5026 (2019)

\bibitem{GFortranDocu}
Documents, F.S.: https://gcc.gnu.org/wiki/GFortranStandards

\bibitem{dolan2022model}
Dolan, H., Rastelli, R.: A model-based approach to assess epidemic risk.
\newblock Statistics in biosciences \textbf{14}(3), 452--484 (2022)

\bibitem{esling2012time}
Esling, P., Agon, C.: Time-series data mining.
\newblock ACM Computing Surveys (CSUR) \textbf{45}(1), 1--34 (2012)

\bibitem{girardi2023seir}
Girardi, P., Gaetan, C.: An seir model with time-varying coefficients for
  analyzing the sars-cov-2 epidemic.
\newblock Risk Analysis \textbf{43}(1), 144--155 (2023)

\bibitem{gleeson2022}
Gleeson, J.P., Brendan~Murphy, T., O’Brien, J.D., Friel, N., Bargary, N.,
  O'Sullivan, D.J.: Calibrating covid-19 susceptible-exposed-infected-removed
  models with time-varying effective contact rates.
\newblock Philosophical Transactions of the Royal Society A \textbf{380}(2214),
  20210120 (2022)

\bibitem{goldberg1989zen}
Goldberg, D.E.: Zen and the art of genetic algorithms.
\newblock In: Proceedings of the 3rd international conference on genetic
  algorithms, pp. 80--85 (1989)

\bibitem{golub1996implementation}
Golub, M.: An implementation of binary and floating point chromosome
  representation in genetic algorithm.
\newblock In: Proceedings of the 18th International Conference on Information
  Technology Interfaces, pp. 417--422 (1996)

\bibitem{guo2021calibrating}
Guo, D., Olesen, J.E., Pullens, J.W., Guo, C., Ma, X.: Calibrating aquacrop
  model using genetic algorithm with multi-objective functions applying
  different weight factors.
\newblock Agronomy Journal \textbf{113}(2), 1420--1438 (2021)

\bibitem{holland1992genetic}
Holland, J.H.: Genetic algorithms.
\newblock Scientific american \textbf{267}(1), 66--73 (1992)

\bibitem{iorio2006incorporating}
Iorio, A.W., Li, X.: Incorporating directional information within a
  differential evolution algorithm for multi-objective optimization.
\newblock In: Proceedings of the 8th annual conference on Genetic and
  evolutionary computation, pp. 691--698 (2006)

\bibitem{klusener2020forecasting}
Kl{\"u}sener, S., Schneider, R., Rosenbaum-Feldbr{\"u}gge, M., Dudel, C.,
  Loichinger, E., Sander, N., Backhaus, A., Del~Fava, E., Esins, J., Fischer,
  M., et~al.: Forecasting intensive care unit demand during the covid-19
  pandemic: A spatial age-structured microsimulation model.
\newblock medRxiv  (2020)

\bibitem{leardi2001genetic}
Leardi, R.: Genetic algorithms in chemometrics and chemistry: a review.
\newblock Journal of Chemometrics: A Journal of the Chemometrics Society
  \textbf{15}(7), 559--569 (2001)

\bibitem{leardi2003nature}
Leardi, R.: Nature-inspired methods in chemometrics: genetic algorithms and
  artificial neural networks.
\newblock Elsevier (2003)

\bibitem{leardi2007genetic}
Leardi, R.: Genetic algorithms in chemistry.
\newblock Journal of Chromatography A \textbf{1158}(1-2), 226--233 (2007)

\bibitem{li1995global}
Li, M.Y., Muldowney, J.S.: Global stability for the seir model in epidemiology.
\newblock Mathematical biosciences \textbf{125}(2), 155--164 (1995)

\bibitem{lucasius1994understanding}
Lucasius, C.B., Kateman, G.: Understanding and using genetic algorithms part 2.
  representation, configuration and hybridization.
\newblock Chemometrics and Intelligent Laboratory Systems \textbf{25}(2),
  99--145 (1994)

\bibitem{monteiro2020influence}
Monteiro, L.H.A., Gandini, D., Schimit, P.H.: The influence of immune
  individuals in disease spread evaluated by cellular automaton and genetic
  algorithm.
\newblock Computer methods and programs in biomedicine \textbf{196}, 105707
  (2020)

\bibitem{ratnavale2022sliding}
Ratnavale, S., Hepp, C., Doerry, E., Mihaljevic, J.R.: A sliding window
  approach to optimize the time-varying parameters of a spatially-explicit and
  stochastic model of covid-19.
\newblock PLOS Global Public Health \textbf{2}(9), e0001058 (2022)

\bibitem{sampson1976adaptation}
Sampson, J.R.: Adaptation in natural and artificial systems (john h. holland)
  (1976)

\bibitem{santos2012comparison}
Santos~Amorim, E.P.d., Xavier, C.R., Campos, R.S., Santos, R.W.d.: Comparison
  between genetic algorithms and differential evolution for solving the history
  matching problem.
\newblock In: International Conference on Computational Science and Its
  Applications, pp. 635--648. Springer (2012)

\bibitem{yu2010introduction}
Yu, X., Gen, M.: Introduction to evolutionary algorithms.
\newblock Springer Science \& Business Media (2010)

\end{thebibliography}

\end{document}